\documentclass[aps,prb,reprint,twocolumn]{revtex4-2}

\usepackage[english]{babel}
\usepackage[utf8]{inputenc}
\usepackage{amsfonts,latexsym,amsthm,amssymb,amsmath,amscd,euscript,graphicx,mathrsfs,cancel,framed,fullpage,hyperref,physics,tensor,braket,simpler-wick}

\makeatletter

\begin{abstract}
The Ising model, often seen as the paradigmatic spin model, has been heavily studied for its mathematical description of ferromagnetism in statistical mechanics. We explore a quantum version of this model, the transverse field Ising model, and investigate how the quantum property of magnetization fluctuates with geometrical frustration, a phenomenon arising from the geometry of a spin system’s ground state. We introduce a general measure for frustration and then implement a computational model built in Python that translates spin lattices into adjacency matrices, solves the spectrum of the transverse field Ising model, calculates the expected magnetization of the ground state, plots the variable over randomly generated spin systems with controllable degrees of geometrical frustration, and compares the evolution of these plots over an increasing transverse field. Our finite-size studies exhibit an expected decrease in magnetization with increasing frustration and suggest the possibility of robust phases in the large N limit over finite ranges of frustration, which may be precursors to spin liquids or spin glasses.
\end{abstract}

\begin{document}

\title{Geometric and bond frustration in transverse field Ising clusters}
\author{Abhiraj Jalagekar}
\address{Amador Valley High School, 1155 Santa Rita Rd, Pleasanton, CA 94566}
\date{\today}
\maketitle

\section{Introduction}


The classical Ising model, often seen as the paradigmatic spin model \cite{ising1925beitrag,pathria2016statistical,mccoy1973two}, is a model of ferromagnetism in statistical mechanics. The transverse field Ising model is a quantum version of the classical model describing qubits with nearest-neighbor couplings in a transverse field, and can be mapped exactly to the classical Ising model in one higher dimension \cite{Sachdev2011Apr}. While this model is paradigmatic in the field of quantum condensed matter theory, it has only been exactly solved in up to two dimensions for the quantum case (up to three dimensions for the classical case) \cite{Ising1925Feb,Onsager1944Feb}. With the advent of quantum simulators with the ability to coherently manipulate tens of qubits and couple them arbitrarily \cite{Altman2021Feb}, the study of the quantum Ising model in various dimensions, on graphs of arbitrary connectivities, and with bond strengths of arbitrary strengths, is worth revisiting.   
\smallskip

Beyond the classical Ising model’s original application in the context of magnetism \cite{RevModPhys.39.883, lenz1920beitrag}, the model’s simplicity has given it a breadth of application in a diversity of settings, as seen through the comprehensive list found in the detailed article by Reinhart and Cuevas \cite{reinhart2022grammar}: a toy model of matter in quantum gravity models \cite{AMBJORN2009251}, lattice gas models \cite{chandler1987introduction}, the Potts model in knot theory \cite{kauffman2001knots}, Hopfield’s proposal in artificial neural networks \cite{hopfield1982neural, PhysRevA.32.1007}, the size of canopy trees in ecology \cite{goodwin2000signs}, a model for flocks of birds \cite{bialek2012statistical}, viruses as quasi-species \cite{anderson1983suggested, tarazona1992error}, and protein folding \cite{bakk2003one, ekeberg2013improved, leuthausser1986exact, leuthausser1987statistical}, just to name a few. No matter the application, it is clearly a model that provides great value to an extensive range of scientific insight.
\break

\begin{flushleft} \textbf{Geometrical Frustration in Spin Systems} \end{flushleft}

In this work, we will be interested in the effects of geometrical frustration in spin systems. First introduced in the context of magnetic systems by Gerard Toulouse in 1977 \cite{vannimenus1977theory, toulouse2008frustration}, the term frustration refers to a phenomenon where qubits in spin systems become \textit{frustrated} due to being forced to be both spin up and down at the same time, thus struggling to pick one spin orientation. More specifically, geometrical frustration is frustration that stems from the lattice geometry of spin configuration. Due to conflicting couplings, a qubit  may face competing interactions that frustrate it. For example, one coupling may favor the qubit being spin up, whereas another coupling may favor the qubit being spin down. Thus, frustration can arise within the spin system depending on how the couplings are arranged geometrically. One simple example of geometrical frustration would be a triangular arrangement of spins with three magnetic ions residing on the three vertices, as studied initially by G. H. Wannier in 1950 \cite{PhysRev.79.357}. The couplings are antiferromagnetic, meaning that the spins between any two vertices must be antiparallel. Considering the first two vertices, we quickly see that one must be spin up and the other spin down for the antiparallel condition to be satisfied. However, the third ion struggles to achieve the lowest energy level along with the other two \cite{PhysRev.79.357, meiri2021cumulative} because it must be antiparallel with both the first two ions. The ion cannot be antiparallel with a spin up ion at the same time that it must be with a spin down ion, and thus, frustration arises within the spin system. This idea of frustration in the triangular lattice can be extended to other more complex lattices. From here on out, we shall reference geometrical frustration as simply frustration for convenience.
\break

\begin{flushleft} \textbf{Graph Theory and Adjacency Matrices} \end{flushleft}

Qubit lattice and coupling geometry can be described by graph theory through what are known as general dimensional graphs. These graphs consist of vertices that represent qubits and edges between vertices that represent couplings between qubits. These edges may be labeled with real numbers where a positive number would represent a ferromagnetic coupling of some magnitude, a negative number would represent an antiferromagnetic coupling of some magnitude, and zero would represent no coupling. These general dimensional graphs can be represented in the language of linear algebra through what are known as adjacency matrices \cite{stevanovic2014spectral}. An $N$ qubit system would be represented by an $N\times N$ adjacency matrix, where the entry in the $i$th row and $j$th column would store the value of the coupling between the $i$th and $j$th qubits. For example, if the 2nd and 3rd qubits had an antiferromagnetic coupling of -100, the entry in the 2nd row and 3rd column of the corresponding adjacency matrix would read -100. Observe that the entry in the 3rd row and 2nd column would also read -100 as the coupling exists equally between both qubits. Extending this logic to the rest of the couplings, we see that adjacency matrices, at least for our purposes, must be symmetric \cite{szabo2015linear, abudayah2021hermitian}. Additionally, the coupling between a qubit and itself is defined to be 0, making the diagonal of all adjacency matrices full of 0’s. To summarize, the conditions for an adjacency matrix, representing spin systems, are first that it must be symmetric and second that its diagonal must contain only 0’s. Adjacency matrices make it easier for computers to interpret general dimensional graphs and perform relevant calculations.
\medskip

\begin{flushleft} \textbf{Outline} \end{flushleft}

The outline of the paper is as follows. In section II, we elaborate upon our research model in greater depth. We describe the primary equations in our calculations, the constructed measures for our variables of interest, and the methods used for answering our research question. We also analyze a simple case of a three qubit system to better illustrate the ideas introduced throughout section II. In section III, we thoroughly investigate our results, the plots generated from our coded simulations, and we answer our research question while also highlighting some interesting observations. In section IV, we conclude with a discussion of our research in a broader context and address potential focuses for future studies covering this topic.
\bigskip

\section{Model}

\begin{flushleft} \textbf{Overview} \end{flushleft}
Our starting point is the Hamiltonian
\begin{equation}\label{eq:H}
    H = -\sum_{i,j} J_{ij} Z_i Z_j - g\sum_i X_i
\end{equation}
which describes a frustrated transverse field Ising model on $N$ qubits $(i =1,\ldots, N)$ with bond strengths $J_{ij}$. When $J_{ij} = J$, this describes the familiar transverse field Ising model. When $J_{ij}$ varies in sign, we say the model is \textit{frustrated}.
\medskip

In this work, we define a measure for frustration through what we call the $s$ parameter, and we study how magnetization over the transverse field Ising model changes for spin systems with different levels of frustration. To do so, we calculate and plot the dependent variable of magnetization against the parameter $s$ (essentially representing frustration) over randomly generated adjacency matrices with varied values of $s$.
\medskip

We define the $s$ parameter to be the fraction of couplings that are negative (antiferromagnetic) in a spin system, or in the context of adjacency matrices, the fraction of negative entries. This parameter ranges from 0, all positive couplings, to 1, all negative couplings. Note that the denominator of this fraction, the total number of entries, excludes the diagonal of 0’s which don’t represent any couplings between distinct qubits, and simply considers the entries above and below the diagonal. The frustration of spin systems is quite approximately measured by this parameter as after all, it is negative couplings that give rise to contradictory conditions for spin alignment and ultimately frustrate qubits. Consider a system of only positive couplings; we quickly see that all the spins must be aligned and that it is completely unfrustrated. Now consider a system of only negative couplings, and we can imagine how various complications could arise from qubits’ spins having to be antiparallel. Therefore, for general purposes, the $s$ parameter is a strong measure of frustration in spin systems.
\break

\begin{flushleft} \textbf{Methods} \end{flushleft}

\medskip


Our goal is to study magnetization vs. frustration, so we produce magnetization vs. $s$ plots to do so. In our code, we control the $s$ value of adjacency matrices by randomly multiplying the also random entries of the matrix by -1 until we have our desired $s$ value. For example, if there were 7 couplings and our $s$ value was 3/7, we would randomly pick 3 couplings whose magnitudes we would negate. We can perform this process for any $s$ value we desire; however, there are important restrictions to be noted. Due to the number of couplings being discrete, $s$ is discrete as well and increases from 0 in increments of 1/${N}\choose{2}$ for an $N\times N$ adjacency matrix. The number ${N}\choose{2}$ is the total number of possible couplings in an $N$ qubit system; thus, increasing the number of negative couplings by 1 would correspond to increasing s, the fraction of negative couplings, by 1/${N}\choose{2}$. For our plots, we generate a certain number of adjacency matrices per each possible $s$ value, $i$/${N}\choose{2}$ where $i$ ranges from $0$ to ${N}\choose{2}$.
\medskip

Next, we calculate the magnetizations of the adjacency matrices by first calculating the Hamiltonians of the spin system according to Eq.\eqref{eq:H}. We do so using Python scientific computing packages like NumPy and SciPy that are capable of performing linear algebra operations such as matrix multiplication and tensor products. We then diagonalize the matrix representing the Hamiltonian to find the smallest eigenvalue algebraically. The eigenvector corresponding to this eigenvalue represents the ground state of our spin system. We then calculate the expectation value of the magnetization of the ground state according to the formula 
\begin{equation}\label{eq:M}
    \overline{M} = \frac{1}{N_\text{couplings}} \sum_{i,j} \langle Z_i Z_j \rangle _{\text{gnd}}
\end{equation}
where $N_{\text{couplings}}$ is the total number of nonzero couplings. 
\medskip

For the final step of answering our research question, we produce plots of magnetization vs. $s$ over numerous qubit graphs with $s$ ranging from 0 to 1. There are several parameters we set when producing a plot. First are the lower and upper bounds for the magnitudes of the entries in our adjacency matrices, which we call $(j, k)$ with $j$ being the lower bound and $k$ being the upper bound. For example, a matrix with $(j, k) = (3, 5)$ would contain entries of real numbers with magnitudes ranging from 3 to 5, inclusive. Recall that their signs are determined randomly by the given $s$ value. The next parameter is simply the number of qubits $N$ in the spin system we model; this parameter plays a role when creating the $N\times N$ adjacency matrices and is simply ubiquitous in our formulae. Note that as $N$ scales up and the spin system grows larger, calculations take increasingly longer to be performed. Throughout the majority of our paper, we stick with $N = 8$, which gives us a sufficiently large enough system to gain insight into the question we wish to answer. The last parameter is the strength of the transverse field (an external magnetic field), or the parameter $g$ as seen in the formula for the Hamiltonian Eq.\eqref{eq:H}. Setting $g$ equal to 0 is the equivalent of turning off the effects of the transverse field. In this work, we initially take a look at results with the transverse field turned off, but later, we analyze interesting trends in the evolution of plots with transverse fields of increasing strengths.
\break

\begin{flushleft} \textbf{Simple Case: Three Qubits} \end{flushleft}

To better introduce the ideas described earlier, we examine a simple case of our model, a three qubit system with the transverse field turned off, and we solve for the ground state and magnetization. Let the three coupling strengths between the qubits be $c_{1}$, $c_{2}$, and $c_{3}$, and without loss of generality, let $c_{1} \le c_{2} \le c_{3}$. Thus, our corresponding adjacency matrix is
$$ J = \left[\begin{array}{ccc}
0 & c_{1} & c_{2}
\\ c_{1} & 0 & c_{3}
\\ c_{2} & c_{3} & 0 \end{array}\right]. $$
Using our initial formula for the Hamiltonian Eq.\eqref{eq:H}, we get the $8\times8$ matrix
\begin{multline*}
    \begin{gathered}
    H = \text{diag}[ (-c_{1}-c_{2}-c_{3}), (-c_{1}+c_{2}+c_{3}), \\ (c_{1}-c_{2}+c_{3}), (c_{1}+c_{2}-c_{3}), (c_{1}+c_{2}-c_{3}), \\ (c_{1}-c_{2}+c_{3}),
    (-c_{1}+c_{2}+c_{3}), (-c_{1}-c_{2}-c_{3})]
    \end{gathered}
\end{multline*}
where diag[ ] represents a diagonal matrix whose entries from left to right and top to bottom on the diagonal are the values inside the brackets. 
Consequently, given that $c_{1} \le c_{2} \le c_{3}$, the ground state energy 
algebraically is one of two options:
\begin{multline*}
    \begin{gathered}
        \textcolor{white}{-----}
        \lambda_{\text{gnd,1}} = (-c_{1}-c_{2}-c_{3}) \\ 
        \textcolor{white}{-----}
        \lambda_{\text{gnd,2}} = (c_{1}+c_{2}-c_{3}).
    \end{gathered}
\end{multline*}
depending on whether $-c_{1}-c_{2} \leq c_{1}+c_{2}$. The corresponding ground states are
\begin{subequations}
    \begin{align*}
    |\lambda\rangle_{\text{gnd,1}} &= \alpha \hat e_1 + \beta \hat e_8\\
|\lambda\rangle_{\text{gnd,2}} &= \gamma \hat e_3 + \delta \hat e_4
    \end{align*}
\end{subequations}
where $|\alpha|^2 +|\beta|^2 = |\gamma|^2 + |\delta|^2 = 1$ and $[\hat e_j]_k = \delta_{jk}$. Thus, we observe that the system has \textit{classical} frustration but lacks quantum frustration without the transverse field due to the differing bond strengths $c_1,c_2,c_3$. 
\newline

Now, we proceed with solving for the $n$ possible expectation values of the ground state’s magnetization. Using its formula from earlier Eq.\eqref{eq:M}, we get
\begin{multline*}
    \begin{gathered}
        \textcolor{white}{---}
        \overline{M_{n}} = \frac{1}{3} \sum_{i,j} \langle Z_i Z_j \rangle _{\text{gnd,$n$}} \\
        \textcolor{white}{--}
        = \frac{1}{3} (|\lambda\rangle_{\text{gnd,$n$}})^\dag (\sum_{i,j} Z_i Z_j) (|\lambda\rangle_{\text{gnd,$n$}}).
    \end{gathered}
\end{multline*}
Performing similar calculations to those of the Hamiltonian except without the coupling strengths, we have
\begin{equation*}
    \sum_{i,j} Z_i Z_j = \text{diag}[3, -1, -1, -1, -1, -1, -1, 3].
\end{equation*}
Substituting this back into the magnetization calculation and multiplying with the ground state vector and its adjoint, we get $\overline{M_1} = 1$ and $\overline{M_2} = -\frac13$.  Recall the condition we had established for choosing between the two ground states: $-c_{1}-c_{2} \leq c_{1}+c_{2}$. This inequality can be rewritten in simpler terms for the purpose of determining the expected magnetization for a three qubit system's ground state:
\begin{multline*}
    \begin{gathered}
        \textcolor{white}{-----}
        c_{1}+c_{2} \geq 0 \text{  }\text{ }\Longrightarrow\text{  }\text{ } \overline{M} = 1 \\
        \textcolor{white}{-----}
        c_{1}+c_{2} < 0 \text{  }\text{ }\Longrightarrow\text{  }\text{ } \overline{M} = -\frac{1}{3}.
    \end{gathered}
\end{multline*}

\begin{figure}
    \centering
{\setlength{\fboxsep}{1pt}
\setlength{\fboxrule}{0.5pt}
\fbox{\includegraphics[width=78.5mm]{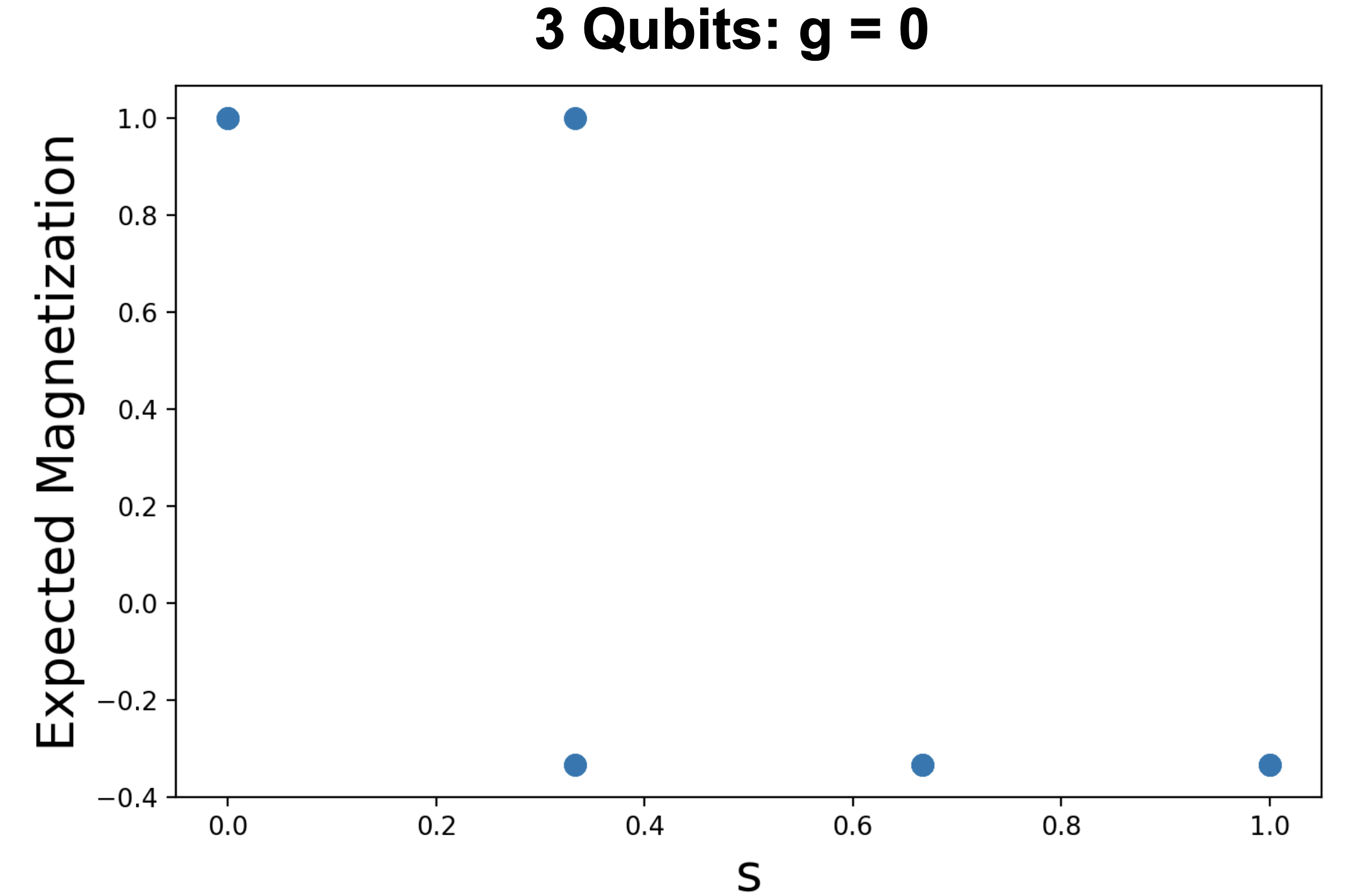}}}
    \caption{Expected magnetization vs. $s$ plot with transverse field turned off for 3 qubits.}
    \label{fig:1}
\end{figure}

Our results for magnetization are confirmed by our code through the plot it generates for three qubits, as seen above in Fig. \ref{fig:1}. For each of the ten systems plotted at each of the four $s$ values, the magnetization is either 1 or $-\frac{1}{3}$, just as we calculated earlier. Furthermore, for matrices with lower $s$ values which indicate mostly positive couplings and would thus satisfy the inequality $c_{1}+c_{2} \geq 0$, our plot shows magnetizations of 1. Likewise, for matrices with higher $s$ values which indicate mostly negative couplings and would thus satisfy the inequality $c_{1}+c_{2} < 0$, our plot shows magnetizations of $-\frac{1}{3}$.
\newline

\section{Results}

\begin{figure}
    \centering
\includegraphics[width=\linewidth]{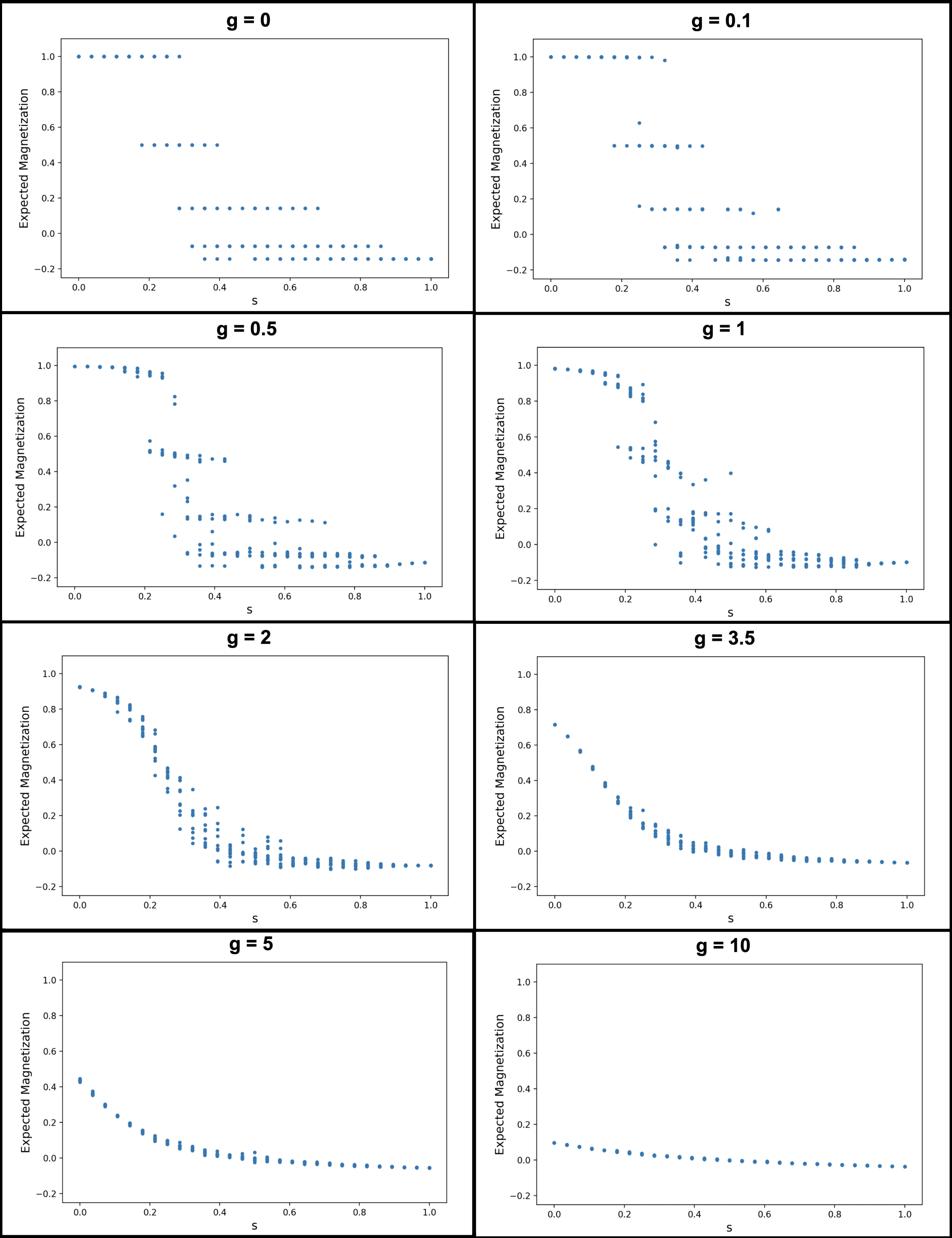}
    \caption{{Evolution of expected magnetization vs. $s$ plots with transverse field strengths increasing from $g$ = 0 to $g$ = 10 for 8 qubits.}}
    \label{fig:2}
\end{figure}

We now proceed to an in-depth investigation of the magnetization of the frustrated transverse field Ising model described in Eq.\eqref{eq:H} as a function of the transverse field and the degree of frustration in the bonds $J_{ij}$. This is plotted in Fig. \ref{fig:2} for various values of transverse field strength increasing from $g$ = 0 to $g$ = 10. Each plot captures an eight-qubit cluster with $(j, k) = (1, 1.1)$ and $s$ values ranging from 0 to 1 in increments of 1/${8} \choose {2}$ $\approx 0.036.$ At each $s$ value, 10 random adjacency matrices are generated, and the corresponding magnetizations and $s$ values are computed and plotted. Observe that it is very possible for these points to overlap, as seen abundantly in the $g$ = 0 plot where some $s$ values give the false impression of having only one point.
\smallskip

Let us make a primary, elementary observation about the behavior of these plots. In general for all $g$, we see that as frustration increases, magnetization tends to decrease. This is expected because an increase in frustration corresponds to a greater fraction of antiferromagnetic bonds, which favor spins anti-aligning, leading to a decrease in the magnetization.  \smallskip 

Next, let us turn our attention to the case of $g$ = 0 plot. In this case, $H$ is diagonal in the $Z$ basis, and the frustration can be considered a \textit{classical} effect because the energy eigenstates are always product states in the $Z$ basis and do not have any entanglement between different qubits. It is remarkable that this classical frustration still has pronounced discrete effects -- magnetization jumps -- which are possible analogs of quantum phase transitions in more complicated frustrated systems. In particular, for these eight qubit clusters, the expected magnetization jumps from $1$ to $0.5$ at a critical value of 
\begin{equation*}
    s_\text{crit} \approx 0.2.
\end{equation*}
From numerical observations, this corresponds to the typical lowest value at which a single row of the adjacency matrix $J_{ij}$ (or column) contains a majority of negative entries, and this results in an order 1 fraction of the cluster having mutually antiferromagnetic couplings. Finally, a sequence of ``classical" phase transitions (i.e. phase transitions between different product states) ensues from tuning $s$ from $0$ to $1$ at critical values $s \approx 0.25, 0.3, 0.35$.
\smallskip

Next, let us review the familiar case of $s=0$ and varying $g$. In this case, the familiar results of the transverse Ising model tell us that the magnetization drops from 1 to 0 with a critical point at $g=1$, which is consistent with our results. \smallskip

Finally, when $s > 0$ and $g > 0$, we observe that the transverse field smears the discrete magnetization jumps which come from the classical frustration. When $g$ is order 1, the states are already far from product states and highly entangled, and we can regard the finite $s$ transition from a polarized to an unpolarized state as a true quantum phase transition or crossover (which has its origin in the classical frustration phase transitions). For instance, when $g=2$, we observe a smooth drop from a polarized to an unpolarized state with a ``critical value" of approximately $s\approx 0.4$. 
\break

\section{Discussion}

In conclusion, we have studied finite-size clusters of a variant of the transverse field Ising model, where we have allowed the bond strengths to vary in magnitude and (crucially) \textit{sign}, allowing for arbitrary degrees of both geometric and bond frustration. We have found that the magnetization of the resulting clusters jumps discretely as we tune the frustration, measured here in a coarse-grained fashion as the fraction $s$ of negative bonds, from $0$ to $1$. These magnetization jumps are due to an order $1$ fraction of the graph belonging to an antiferromagnetic sub-graph. With the addition of the transverse field, these transitions are broadened and smoothed.
\smallskip

Although this study has been done with a finite system size ($N = 8$), we include a few predictions about the behavior of magnetization with frustration in larger systems sizes $(N\to \infty$). In this case, we expect that frustration is tunable as a continuous parameter (as the fraction of negative bonds) and the crossovers at finite $g$ from polarized to unpolarized states may be sharpened into true phase transitions. The classical frustration case ($g=0$) may be amenable analytically. Whether these expectations are borne out remains an interesting open question.
\smallskip

Finally, the study of Ising clusters is a cartoon model of the transverse field Ising model in arbitrary dimensions because, for instance, the coordination number of a cubic lattice in $d$ dimensions grows as $2^d$, and this behavior is mimicked by an all-to-all coupled graph. In future studies, it will be interesting to understand which features of the general $d$ dimensional transverse field Ising model are well-captured by randomly coupled all-to-all Ising clusters.

\bibliographystyle{unsrt}
\bibliography{references}

\begin{thebibliography}{10}

\bibitem{ising1925beitrag}
E.~Ising.
\newblock Beitrag zur theorie des ferromagnetismus.
\newblock {\em Z. Phys.}, 31:253--258, 1925.

\bibitem{pathria2016statistical}
Raj~Kumar Pathria.
\newblock {\em Statistical mechanics}.
\newblock Elsevier, 2016.

\bibitem{mccoy1973two}
Barry~M McCoy and Tai~Tsun Wu.
\newblock {\em The two-dimensional Ising model}.
\newblock Harvard University Press, 1973.

\bibitem{Sachdev2011Apr}
Subir Sachdev.
\newblock {\em {Quantum Phase Transitions}}.
\newblock Cambridge University Press, Cambridge, England, UK, April 2011.

\bibitem{Ising1925Feb}
Ernst Ising.
\newblock {Beitrag zur Theorie des Ferromagnetismus}.
\newblock {\em Z. Phys.}, 31(1):253--258, February 1925.

\bibitem{Onsager1944Feb}
Lars Onsager.
\newblock {Crystal Statistics. I. A Two-Dimensional Model with an Order-Disorder Transition}.
\newblock {\em Phys. Rev.}, 65(3-4):117--149, February 1944.

\bibitem{Altman2021Feb}
Ehud Altman, Kenneth~R Brown, Giuseppe Carleo, Lincoln~D Carr, Eugene Demler, Cheng Chin, Brian DeMarco, Sophia~E Economou, Mark~A Eriksson, Kai-Mei~C Fu, et~al.
\newblock Quantum simulators: Architectures and opportunities.
\newblock {\em PRX Quantum}, 2(1):017003, 2021.

\bibitem{RevModPhys.39.883}
STEPHEN~G. BRUSH.
\newblock History of the lenz-ising model.
\newblock {\em Rev. Mod. Phys.}, 39:883--893, Oct 1967.

\bibitem{lenz1920beitrag}
Wilhelm Lenz.
\newblock Beitrag zum verst{\"a}ndnis der magnetischen erscheinungen in festen k{\"o}rpern.
\newblock {\em Z. Phys.}, 21:613--615, 1920.

\bibitem{reinhart2022grammar}
Tobias Reinhart and Gemma De~las Cuevas.
\newblock The grammar of the ising model: A new complexity hierarchy.
\newblock {\em arXiv preprint arXiv:2208.08301}, 2022.

\bibitem{AMBJORN2009251}
J.A. Ambjørn, K.N. Anagnostopoulos, R.~Loll, and I.~Pushkina.
\newblock Shaken, but not stirred—potts model coupled to quantum gravity, 2009.

\bibitem{chandler1987introduction}
David Chandler.
\newblock Introduction to modern statistical.
\newblock {\em Mechanics. Oxford University Press, Oxford, UK}, 5:449, 1987.

\bibitem{kauffman2001knots}
Louis~H Kauffman.
\newblock {\em Knots and physics}, volume~1.
\newblock World scientific, 2001.

\bibitem{hopfield1982neural}
John~J Hopfield.
\newblock Neural networks and physical systems with emergent collective computational abilities.
\newblock {\em Proceedings of the national academy of sciences}, 79(8):2554--2558, 1982.

\bibitem{PhysRevA.32.1007}
Daniel~J. Amit, Hanoch Gutfreund, and H.~Sompolinsky.
\newblock Spin-glass models of neural networks.
\newblock {\em Phys. Rev. A}, 32:1007--1018, Aug 1985.

\bibitem{goodwin2000signs}
Brian~C Goodwin.
\newblock {\em Signs of life: How complexity pervades biology}.
\newblock Basic Books, 2000.

\bibitem{bialek2012statistical}
William Bialek, Andrea Cavagna, Irene Giardina, Thierry Mora, Edmondo Silvestri, Massimiliano Viale, and Aleksandra~M Walczak.
\newblock Statistical mechanics for natural flocks of birds.
\newblock {\em Proceedings of the National Academy of Sciences}, 109(13):4786--4791, 2012.

\bibitem{anderson1983suggested}
Philip~W Anderson.
\newblock Suggested model for prebiotic evolution: the use of chaos.
\newblock {\em Proceedings of the National Academy of Sciences}, 80(11):3386--3390, 1983.

\bibitem{tarazona1992error}
Pedro Tarazona.
\newblock Error thresholds for molecular quasispecies as phase transitions: From simple landscapes to spin-glass models.
\newblock {\em Physical Review A}, 45(8):6038, 1992.

\bibitem{bakk2003one}
Audun Bakk and Johan~S H{\o}ye.
\newblock One-dimensional ising model applied to protein folding.
\newblock {\em Physica A: Statistical Mechanics and its Applications}, 323:504--518, 2003.

\bibitem{ekeberg2013improved}
Magnus Ekeberg, Cecilia L{\"o}vkvist, Yueheng Lan, Martin Weigt, and Erik Aurell.
\newblock Improved contact prediction in proteins: using pseudolikelihoods to infer potts models.
\newblock {\em Physical Review E}, 87(1):012707, 2013.

\bibitem{leuthausser1986exact}
Ira Leuth{\"a}usser.
\newblock An exact correspondence between eigen’s evolution model and a two-dimensional ising system.
\newblock {\em The Journal of chemical physics}, 84(3):1884--1885, 1986.

\bibitem{leuthausser1987statistical}
Ira Leuth{\"a}usser.
\newblock Statistical mechanics of eigen's evolution model.
\newblock {\em Journal of statistical physics}, 48:343--360, 1987.

\bibitem{vannimenus1977theory}
J~Vannimenus and G~Toulouse.
\newblock Theory of the frustration effect. ii. ising spins on a square lattice.
\newblock {\em Journal of Physics C: Solid State Physics}, 10(18):L537, 1977.

\bibitem{toulouse2008frustration}
G{\'e}rard Toulouse.
\newblock The frustration model.
\newblock In {\em Modern Trends in the Theory of Condensed Matter: Proceedings of the XVI Karpacz Winter School of Theoretical Physics, February 19--March 3, 1979 Karpacz, Poland}, pages 195--203. Springer, 2008.

\bibitem{PhysRev.79.357}
G.~H. Wannier.
\newblock Antiferromagnetism. the triangular ising net.
\newblock {\em Phys. Rev.}, 79:357--364, Jul 1950.

\bibitem{meiri2021cumulative}
Snir Meiri and Efi Efrati.
\newblock Cumulative geometric frustration in physical assemblies.
\newblock {\em Physical Review E}, 104(5):054601, 2021.

\bibitem{stevanovic2014spectral}
Dragan Stevanovic.
\newblock {\em Spectral radius of graphs}.
\newblock Academic Press, 2014.

\bibitem{szabo2015linear}
Fred Szabo.
\newblock {\em The linear algebra survival guide: illustrated with Mathematica}.
\newblock Academic Press, 2015.

\bibitem{abudayah2021hermitian}
Mohammad Abudayah, Omar Alomari, and Torsten Sander.
\newblock Hermitian adjacency matrices of mixed graphs.
\newblock {\em arXiv preprint arXiv:2103.16969}, 2021.

\end{thebibliography}

\end{document}